\documentclass[amsmath,amssymb,twocolumn]{revtex4}
\usepackage{graphicx}
\usepackage{dcolumn}
\usepackage{bm}
\usepackage{amsmath,mathrsfs}
\usepackage{graphicx,epsfig}
\usepackage{subfigure}
\usepackage{hyperref}
\usepackage{epsf}
\usepackage{color}

\begin{document}
%
%
\title{{\bf Wormhole geometries and energy conditions in linear and non-linear \(f(Q,L_m)\) gravity}}
\author{S. Rastgoo}\email{rastgoo@sirjantech.ac.ir}
\author{ F. Parsaei }\email{fparsaei@gmail.com}
%
%
\affiliation{ Physics Department , Sirjan University of Technology, Sirjan 78137, Iran.}
%
%




\begin{abstract}
\par
This study explores static and spherically symmetric wormhole configurations in an extended formulation of symmetric teleparallel gravity governed by $f(Q,L_m)=\alpha (-Q)+\beta L_m $ and  $f(Q,L_m)= -Q+L_m^{\alpha} $ (where $Q$ is the non-metricity scalar, $L_m$ is the matter Lagrangian, and $\alpha, \beta$ are constants). By varying the action with respect to the metric, we obtain the corresponding field equations. We conclude that the linear $f(Q,L_m)=\alpha (-Q)+\beta L_m$ framework successfully supports physically realistic wormholes without requiring extreme energy condition violations. It has been demonstrated that spacetime solutions violating both the radial and lateral null energy conditions  in general relativity can nevertheless satisfy the NEC in $f(Q, L_m)$ gravity. Within non-linear $f(Q,L_m)$ gravity, non-exotic asymptotically flat wormhole solutions are derived using linear and asymptotically linear equations of state. Our results open new avenues to link modified gravity models with the observable astrophysical traits of exotic spacetimes.\\
\end{abstract}

\maketitle

\section{Introduction}

A wormhole is a theoretical tunnel connecting two far-off regions of spacetime, which could theoretically allow for shortcuts across the universe. Derived from Einstein’s theory of General Relativity (GR) and related frameworks, wormholes remain among the most fascinating predictions in modern physics \cite{WH}. First formally introduced within the framework of GR in 1935, wormholes have since become a central concept in the theoretical study of spacetime travel \cite{Visser}.To be traversable in both directions, a wormhole must feature a "flared-out" throat. In the framework of GR, achieving this flared-out geometry requires a violation of the Null Energy Condition (NEC)\cite{Visser}.

The exploration of accelerated cosmic expansion  provides a compelling framework for investigating wormhole geometries supported by phantom matter \cite{ phantom2, phantom1}. In an effort to reduce the reliance on exotic matter, some researchers have investigated various wormhole geometries, such as thin-shell wormholes \cite{cut, cut2, cut3}, models with a variable equation of state (EoS) \cite{Remo, variable}, and those described by a polynomial EoS \cite{foad}.

Over the past two decades, modified theories of gravity have gained significant prominence in addressing unresolved cosmological and quantum challenges, notably the universe's accelerated expansion, the nature of dark matter and dark energy, and the quantization of gravity on curved manifolds. By incorporating additional geometric or physical structures, modified gravity theories frequently enable the construction of traversable wormholes without the need for exotic matter. Notable examples explored in the literature include wormholes within Braneworld models \cite{b1, b2, b3}, $f(Q)$ gravity \cite{fq, fq2, fq4, fq44}, $f(R)$ gravity \cite{Nojiri, fR11,  fR55}, Einstein-Cartan gravity \cite{Cartan, Cartan1, Cartan2, Cartan3}, Rastall gravity \cite{Rast, Rast1}, Rastall–Rainbow gravity \cite{RaR, RaR1}, $f(T,\mathcal{T})$ gravity \cite{Must1, Err, Riz, Must2, foad3}, $f(R, L_m)$ gravity \cite{L1, L2, L7, L10, L12, SR2}, $f(R,T)$ gravity \cite{Moa, Zub, fr2, Sha, Ban, Sarkar, SR1, foad4} and $f(Q,T)$ gravity \cite{5, 6, 7, 9, 11, 12a, 1, 12, 2, 3, 4, 8, 13, 10, 14, fq-sara, Sodab, Sara, 15, 16}.

While GR traditionally explains gravity through spacetime curvature—quantified by the Riemann tensor—equivalent geometric descriptions also exist. These include the Teleparallel Equivalent of General Relativity (TEGR), which attributes gravity to torsion, and the Symmetric Teleparallel Equivalent of General Relativity (STEGR), where non-metricity governs the gravitational field \cite{Hei}. A key generalization of the teleparallel framework involves introducing an explicit non-minimal coupling between geometry and matter. This development gives rise to theories like $f(Q, T)$ gravity \cite{fqt} and, more fundamentally, $f(Q, L_m)$ gravity \cite{fql}, in which the gravitational Lagrangian is a function of both the non-metricity scalar $Q$ and the matter Lagrangian density $L_m$. Cosmological phenomena within the framework of $f(Q,L_m)$ gravity have been widely investigated by researchers \cite{fql1, fql2, fql3, fql4}. Lagrangian-based modified gravity theories, notably $f(Q,L_m)$ gravity, offer a compelling geometric framework for explaining cosmic acceleration and self-gravitating systems without relying on exotic dark matter or energy. By generalizing the Einstein-Hilbert action with a Lagrangian density $\mathcal{L}_g = f(Q,L_m)/\left(16\pi G\right)$ that couples the Ricci scalar $R$ with the trace of the energy-momentum tensor (EMT) $T$, these models introduce non-minimal geometry-matter interactions into the field equations. These extra geometric terms naturally alter the Raychaudhuri equations and the effective equation of state, presenting a viable framework for sustaining traversable wormholes, resolving galactic rotation curve anomalies, and investigating novel violations of energy conditions (ECs) in astrophysics.

Recently, Suhasini et al. investigated wormhole solutions within the framework of $f(Q,L_m)=-\alpha Q+ 2 L_m+ \beta$ gravity \cite{Suha}. In \cite{Man}, the study explores the possibility of traversable wormhole formation and their thermodynamic properties under the $f(Q, L_m, T)$ gravity model. Given that wormholes remain largely unexplored in this specific gravity framework, we are motivated to further investigate wormhole geometry in $f(Q, L_m)$ gravity.

This manuscript is structured as follows: Section \ref{sec2} presents the theoretical framework of wormholes and $f(Q,L_m)$ gravity alongside the classical ECs. In Section \ref{sec3}, we examine specific solutions that satisfy these conditions, while Section \ref{sec4} outlines our principal findings. Unless stated otherwise, we adopt gravitational units with $c = 8\pi G = 1$.

\section{Basic formulation of wormhole and \texorpdfstring{$f(Q,L_m)$}{f(Q,Lm)}  gravity} \label{sec2}
In this paper, we investigate wormhole solutions within the framework of the following spherically symmetric spacetime:
\begin{equation}\label{1}
ds^2=-U(r)dt^2+\frac{dr^2}{1-\frac{b(r)}{r}}+r^2(d\theta^2+\sin^2\theta
d\phi^2),
\end{equation}
where $U(r)=\exp2\Phi(r)$. In this framework, $\Phi(r)$ acts as the redshift function—responsible for gravitational redshift—and $b(r)$ serves as the shape function, which determines the wormhole's spatial geometry. The wormhole's throat functions as a passage connecting either two different universes or two distant locations within a single universe, and is governed by the condition
\begin{equation}\label{2}
b(r_0)=r_0,
\end{equation}
where $r_{0}$ is the throat of wormhole. For traversability, the shape function must satisfy the flaring-out condition
\begin{equation}\label{3}
b'(r_0)<1,
\end{equation}
while the absence of event horizons is guaranteed by
\begin{equation}\label{4}
b(r)<r,\ \ {\rm for} \ \ r>r_0.
\end{equation}
In addition, any viable wormhole solution must be asymptotically flat so that it matches the observed large-scale behavior of the universe,
\begin{equation}\label{5}
\lim_{r\rightarrow \infty}\frac{b(r)}{r}=0,\qquad   \lim_{r\rightarrow \infty}U(r)=1.
\end{equation}
In various modified theories of gravity, choosing a constant redshift function significantly simplifies the field equations for wormhole solutions. Here, we employ the standard assumption of a vanishing redshift function. The system is modeled as an anisotropic fluid characterized by the energy-momentum tensor(EMT) $T^{\mu}_{\nu} = \text{diag}[-\rho, p, p_t, p_t]$, with $\rho$, $p$, and $p_t$ denoting the energy density, radial pressure, and transverse pressure, respectively. We now present a concise overview of $f(Q,L_m)$ gravity.

The action for the $f(Q,L_m)$ theory of gravity  is given by
\begin{equation}\label{6a}
S=\int\frac{1}{2}\,f(Q,L_m)\sqrt{-g}\,d^4x,
\end{equation}
where $f(Q, L_m)$ represents an arbitrary function of the non-metricity scalar $Q$ and the matter Lagrangian density $L_m$. Here, $\sqrt{-g}$ is the square root of the determinant of the metric tensor, \cite{fql}.The non-metricity tensor is defined as \cite{6}
\begin{equation}\label{6ab}
Q_{\lambda\mu\nu} = \nabla_{\lambda} g_{\mu\nu}.
\end{equation}
Furthermore, the superpotential (alternatively known as the non-metricity conjugate) is given by
\begin{eqnarray}\label{6ab1}
P^{\alpha}{}_{\mu\nu} = \frac{1}{4} \Big[ &-& Q^{\alpha}{}_{\mu\nu} + 2Q_{(\mu}{}^{\alpha}{}_{\nu)} \nonumber \\
&+& Q^{\alpha} g_{\mu\nu} - \tilde{Q}^{\alpha} g_{\mu\nu} - \delta^{\alpha}_{(\mu} Q_{\nu)} \Big],
\end{eqnarray}
where
\begin{equation}\label{a1}
Q_{\alpha} = Q_{\alpha}{}^{\mu}{}_{\mu}, \quad \tilde{Q}_{\alpha} = Q^{\mu}{}_{\alpha\mu}
\end{equation}
represent the two independent traces of the non-metricity tensor. The non-metricity scalar is then defined by \cite{Suha}
\begin{eqnarray}\label{a2}
Q &=& -Q_{\alpha\mu\nu} P^{\alpha\mu\nu} \nonumber \\
&=& -g^{\mu\nu} \left( L^{\beta}{}_{\alpha\mu} L^{\alpha}{}_{\nu\beta} - L^{\beta}{}_{\alpha\beta} L^{\alpha}{}_{\mu\nu} \right),
\end{eqnarray}
with the distortion tensor $L^{\beta}{}_{\mu\nu}$ given by
\begin{equation}\label{a3}
L^{\beta}{}_{\mu\nu} = \frac{1}{2} Q^{\beta}{}_{\mu\nu} - Q_{(\mu}{}^{\beta}{}_{\nu)}.
\end{equation}
Finally, for the metric specified in (\ref{1}), the non-metricity scalar takes the explicit form:
\begin{equation}\label{aa3}
Q = -\frac{b}{r^2} \left[ \frac{r b' - b}{r(r-b)} \right].
\end{equation}
The field equations of $f(Q,T)$ gravity are subsequently obtained by applying the variational principle to the action given in (\ref{6a}):
 \begin{multline}\label{6b}
\frac{2}{\sqrt{-g}}\bigtriangledown_\alpha\left(\sqrt{-g}\,f_Q\,P^\alpha\;_{\mu\nu}\right)+\frac{1}{2}g_{\mu\nu}f \\
+f_Q\left(P_{\mu\alpha\beta}\,Q_\nu\;^{\alpha\beta}-2\,Q^
{\alpha\beta}\,\,_{\mu}\,P_{\alpha\beta\nu}\right)=\frac{1}{2}f_{L_m}\left(L_m g_{\mu\nu}- T_{\mu\nu}\right),
\end{multline}
where $f_{Q}=\frac{\partial f}{\partial Q}$ and $f_{L_m}=\frac{\partial f}{\partial L_m}$.
One can write the EMT  as
 \begin{equation}\label{a4}
T_{\mu\nu}=-\frac{2}{\sqrt{-g}}\frac{\delta\left(\sqrt{-g}\,L_m\right)}{\delta g^{\mu\nu}},
\end{equation}
It is assumed that $L_m$ depends solely on the metric and not on its derivatives.

The specific field equations depend on the chosen functions $f(Q,L_m)$ and the matter Lagrangian $L_m$. In theories of this type, the matter Lagrangian is typically set to one of three standard forms: $L_{m} = -\rho$, $L_{m} = T$, or $L_{m} = P$. Here, the trace of the energy-momentum tensor is given by $T = -\rho + p + 2p_{t}$, and the average pressure is $P = \frac{p + 2p_{t}}{3}$. To construct wormhole solutions in the subsequent section, we adopt the choice $L_{m} = -\rho$ alongside  linear and  non-linear $f(Q,L_m)$ models.

Because the math allows for almost any distribution of matter and energy to create spacetime curvature, physicists needed a set of rules to prevent physically absurd scenarios (like infinite densities or negative mass everywhere). These rules are called ECs. These conditions are mathematical inequalities applied to the stress-energy tensor ($T_{\mu\nu}$) that ensure the theory respects common-sense physical principles, such as:
\begin{itemize}
  \item Gravity should generally be attractive.
  \item Energy density cannot be negative for a normal observer.
  \item Matter and energy cannot travel faster than the speed of light.
\end{itemize}
The most prominent energy conditions include:
\begin{itemize}
  \item The Weak Energy Condition (WEC): States that any observer will measure a non-negative energy density. (Mathematically: $T_{\mu\nu}u^\mu u^\nu \ge 0$ for any timelike vector $u^\mu$).
  \item The Null Energy Condition (NEC): A slightly more relaxed condition stating that energy density is non-negative along null (light-like) trajectories. ($T_{\mu\nu}k^\mu k^\nu \ge 0$).
  \item The Dominant Energy Condition (DEC): States that mass-energy cannot travel faster than light.
   \item The Strong Energy Condition (SEC): States that gravity must always be attractive (focusing matter together).
\end{itemize}
Research shows that wormhole solutions in GR break standard ECs. The fundamental types of energy conditions are:
\begin{eqnarray}\label{21}
\textbf{NEC}&:& \rho+p\geq 0,\quad \rho+p_t\geq 0 \\
\label{21a}
\textbf{WEC}&:& \rho\geq 0, \rho+p\geq 0,\quad \rho+p_t\geq 0, \\
\textbf{DEC}&:& \rho\geq 0, \rho-|p|\geq 0,\quad \rho-|p_t|\geq 0, \\
\textbf{SEC}&:& \rho+p\geq 0,\, \rho+p_t\geq 0,\rho+p+2p_t \geq 0. \label{21b}
\end{eqnarray}
Although these conditions are essential for classic results like the Hawking–Penrose singularity theorems, black‑hole area theorems, and mass positivity, quantum phenomena (e.g., the Casimir effect) and certain cosmological models (e.g., dark energy where $\omega < -1$) are known to violate them. This limitation has driven the development of more flexible averaged and quantum energy conditions. As discussed in \cite{fq}, we will analyze the ECs by introducing
\begin{eqnarray}\label{22}
 H(r)&=& \rho+p ,\, H_1(r)= \rho+p_t,\, H_2(r)= \rho-|p|, \nonumber \\
 H_3(r)&=&\rho-|p_t|,\, H_4(r)= \rho+p+2p_t .
\end{eqnarray}


\section{Non-exotic wormhole solutions }\label{sec3}
In this section, we investigate non-exotic wormhole solutions using two distinct $f(Q,L_m)$ models: the first assumes a linear dependence on $L_m$, while the second incorporates a nonlinear formulation. Utilizing the field equations in (\ref{6b}) along with metric (\ref{1}) and Eqs.(\ref{6a}-\ref{a4}), one can reach  the following field equation
 \begin{equation}\label{19aa}
f_{Q}(-\frac{b^{\prime}}{r^{2}})+\frac{1}{2}(f-Qf_{Q})-b\frac{f^{\prime}_{Q}}{r^{2}}=0,
\end{equation}

\begin{equation}\label{11cc}
f_{Q}(-\frac{b}{r^{3}})+\frac{1}{2}(f-Qf_{Q})-b\frac{f^{\prime}_{Q}}{r^{2}}=-\frac{1}{2}f_{L_{m}}(\rho+p_{r}),
\end{equation}

\begin{equation}\label{11ccc}
f_{Q}(\frac{b-rb^{\prime}}{2r^{3}})+\frac{1}{2}(f-Qf_{Q})=-\frac{1}{2}f_{L_{m}}(\rho+p_{t}).
\end{equation}
Using a specific form of $f(Q,L_m)$ and a linear EoS, we derive exact, asymptotically flat wormhole solutions.

\subsection{\texorpdfstring{$f(Q,L_m)=- \alpha Q+\beta L_m$}{f(Q,Lm)=-  Q+ Lm} }\label{sec31}
As the first example, we assume a linear function for $f(Q,L_{m})$ as follows
\begin{equation}\label{99a}
f(Q, L_m)=-\alpha  Q+\beta L_m,
\end{equation}
where $\alpha$ and $\beta$ are free parameters. The linear function is the simplest and most straightforward choice in Lagrangian-based gravity theories.
Using Eq.(\ref{99a}) in Eqs.(\ref{19aa}-\ref{11cc}) gives
\begin{equation}\label{3e}
  \rho =2(\frac{\alpha}{\beta})\frac{b^{\prime}}{r^{2}},
\end{equation}
\begin{equation}\label{3f}
 p_{r} =-2(\frac{\alpha}{\beta})\frac{b}{r^{3}},
\end{equation}
\begin{equation}\label{3g}
 p_{t} =2(\frac{\alpha}{\beta})\frac{b-rb^{\prime}}{2r^{3}},
\end{equation}
It is straightforward to show that
\begin{equation}\label{3g1}
 \rho+p_{r} =2(\frac{\alpha}{\beta})\frac{rb^{\prime}-b}{r^{3}},
\end{equation}
and
\begin{equation}\label{3g2}
 \rho+p_{t} =2(\frac{\alpha}{\beta})\frac{rb^{\prime}+b}{2r^{3}},
\end{equation}
These equations indicates a relation between modified theory $f(Q,L_m)$ with GR via coupling constant $2\frac{\alpha}{\beta}$. It is clear that the result of this model is analogous to linear $f(R,T)$ theory \cite{SR1}, linear $f(T,\tau)$ \cite{foad3}, linear $f(Q,T)$ theory \cite{fq-sara} and linear $f(R,L_m)$ theory \cite{SR2} . Therefore, the wormhole solution which contravene both radial and lateral NEC in the context of the GR may still satisfy the NEC in the framework of $f(Q,L_m)$ gravity when the condition
\begin{equation}\label{s1}
\frac{\alpha}{\beta}<0
\end{equation}
is fulfilled. One can easily show that using a linear EoS
\begin{equation}\label{s2}
p_{r}=\omega\rho
\end{equation}
in the Eqs.(\ref{3e}-\ref{3f}) yields
\begin{equation}\label{s3}
b=r^{m}
\end{equation}
where
\begin{equation}\label{s4}
m=-\frac{1}{\omega}
\end{equation}
This well-known shape function in the literature, satisfies all necessary conditions for the existence of the traversable wormhole solution. Also, the asymptotically flat requirement meets as long as the condition $m<1$ continues to hold. The positive energy density results in when $m<0$. So
\begin{equation}\label{s5}
\omega>0
\end{equation}
As discussed above, it is evident from field equations (\ref{3e}-\ref{3g}) that the linear model of $f(Q,L_m)$ modifies the gravitational coupling constant as follows
\begin{equation}\label{s6}
G^{f}=2\frac{\alpha}{\beta}G^{GR},
\end{equation}
where $G^{f}$ and $G^{GR}$ infer the gravitational coupling constant in the framework of modified $f(Q,L_m)$ theory and GR, respectively. This result indicates that the solutions analyzed in \cite{SR1, foad3, fq-sara} can be considered as non-exotic wormhole solutions in the context of $f(Q,L_m)$ theory. For instance, we consider
\begin{equation}\label{s7}
b(r)=(\omega_\infty +D)^{1/n\omega_\infty}(\omega_\infty r^n+D)^{-1/n\omega_\infty},
\end{equation}
This shape function was earlier studied in \cite{SR1}, as a wormhole solution with variable EoS
\begin{equation}\label{b10}
p(r)=\left(\omega_\infty+\frac{D}{r^n}\right)\rho(r)
 \end{equation}
where $\omega_\infty$ is the EoS parameter at large scales and $D$ is a constant. Substituting the shape function (\ref{s7}) in the field equations (\ref{3e}-\ref{3g}), gives
\begin{equation}\label{c2}
\rho(r)=-2\frac{\alpha}{\beta}(\omega_\infty+D)^{\frac{1}{n\omega_\infty}}\frac{\left(\omega_\infty r^n+D\right)^{-\frac{1}{n\omega_\infty}-1}}{r^{3-n}}.
\end{equation}
When $\frac{\alpha}{\beta} < 0$, it is straightforward to demonstrate that asymptotically flat wormhole solutions devoid of exotic matter can be sustained, provided the following conditions are met \cite{SR1}
\begin{eqnarray}\label{29c1}
0\leq \omega_\infty\leq 1-D, \\
\omega_\infty>-D.
\end{eqnarray}
As previously stated, the solutions which breach NEC in GR may still meet NEC in $f(Q,L_m)$ gravity.

\subsection{\texorpdfstring{$f(Q,L_m)=- Q+L_m^\alpha$}{f(Q,Lm)=- Q+Lm} }\label{sec32}
The non-linear function
 \begin{equation}\label{s8}
f(Q,L_m)=-Q+\beta L_{m}^{\alpha}
\end{equation}
is the next case to explore the non-exotic wormhole solution. This model can be regarded as an interaction between matter and gravitational fields through the metric tensor.
 This interaction reveals how matter influences the spacetime curvature. In the case $\alpha=1$, GR is attainable. Using Eq.(\ref{s8}), the field equations (\ref{19aa}-\ref{11ccc}) give
 \begin{equation}\label{s9}
-(\frac{2}{\beta})\frac{b^{\prime}}{r^{2}}=(-\rho)^{\alpha}
\end{equation}
 \begin{equation}\label{s10}
p=\rho \bigg[-\frac{b}{\alpha rb^{\prime}}+\frac{1}{\alpha}-1 \bigg]
\end{equation}
 \begin{equation}\label{s11}
p_{t}=\rho \bigg[\frac{1}{\alpha}\bigg(\frac{b-rb^{\prime}}{2rb^{\prime}}\bigg)+\frac{1}{\alpha}-1\bigg]
\end{equation}
Numerous strategies exist for finding wormhole solutions, with the most conventional approach relying on the choice of an EoS. In this work, we consider linear EoSs
\begin{eqnarray}\label{s12}
p=\omega_{r}\rho \nonumber \\
p_{t}=\omega_{t}\rho
\end{eqnarray}
where $\omega_{r}$ and $\omega_{t}$ are the EoS parameters. In according to Eq.(\ref{s10}), one can show that
\begin{equation}\label{s13}
\omega_{r}=-\frac{b}{\alpha rb^{\prime}}+\frac{1}{\alpha}-1
\end{equation}
which leads to
\begin{equation}\label{s14}
b=r^{m}
\end{equation}
where
\begin{equation}\label{s15}
m=\frac{1}{\alpha\omega_{r}+\alpha-1}
\end{equation}
As the most prominent wormhole solution in literature, the power-law shape function successfully satisfies all necessary conditions for wormhole construction. Using the shape function (\ref{s14}) in (\ref{s11}) result in
\begin{equation}\label{s16}
\omega_{t}=\frac{1}{2m\alpha}+\frac{1}{2\alpha}-1
\end{equation}
It is easy to show that
\begin{equation}\label{s17}
\rho=-(-\frac{2m}{\beta})^{\frac{1}{\alpha}}r^{\frac{m-3}{\alpha}}
\end{equation}
In order to have positive energy density, the condition
\begin{equation}\label{s18}
(-\frac{2m}{\beta})^{\frac{1}{\alpha}}<0
\end{equation}
must be satisfied. For $\alpha = \frac{1}{2n+1}$ (where $n$ is a positive integer),  the condition (\ref{s18})  is satisfied in the following two situations
\begin{eqnarray}\label{s19}
\textbf{case 1} &:& 0<m<1 \;, \;\;\;\beta>0 \nonumber\\
\textbf{case 2} &:& m<0 \;, \;\;\;\;\;\beta<0 . \nonumber\\
\end{eqnarray}
In the next step, we analyze the ECs. The condition $H\geq0$ along with positivity of the energy density lead to $\omega_{r}\geq-1$. By using Eq.(\ref{s15}), we obtain
\begin{equation}\label{s20}
\frac{1}{m}+1=\alpha(\omega_{r}+1)
\end{equation}
Since $\alpha$ is a positive integer and $\omega_r + 1$ must be positive, it follows that
\begin{equation}\label{s21}
\frac{1}{m}>-1
\end{equation}
This condition satisfies the \textbf{case1} while for \textbf{case2} just in the range $m<-1$ is acceptable. Also, the positivity of the energy density with $H_{1}>0$ result in $\omega_{t}>-1$ which according to Eq.(\ref{s16}), meet the condition (\ref{s21}). So, NEC and WEC for the \textbf{case1} is straightforward while for the \textbf{case2} is reliable in the range $m<-1$ .
 To investigate the SEC, the condition $H_{4}>0$ will be studied. Utilizing Eqs. (\ref{s20}) and (\ref{s16}) in the condition $H_{4}>0$ leads to
\begin{equation}\label{s22}
1+\frac{1}{m}>\alpha.
\end{equation}
 To examine the DEC, the condition $H_{2}>0$ leads to $1>|\omega_{r}|$. Using Eq.(\ref{s20}), the value of $\frac{1}{\alpha}(\frac{1}{m}+1)$ determines the validity of $1>|\omega_{r}|$. Firstly, we suppose $\omega_{r}<0$ which results in
 \begin{equation}\label{s23a}
\frac{1}{\alpha}(\frac{1}{m}+1)<1.
\end{equation}
For $\omega_{r}<0$,
\begin{equation}\label{s23b}
H_2(r)=1-|\omega_r|=\frac{1}{\alpha}(\frac{1}{m}+1)
\end{equation}
so
\begin{equation}\label{s23}
\frac{1}{\alpha}(\frac{1}{m}+1)>0
\end{equation}
must hold. Since $\alpha$ assumed to be positive,
\begin{equation}\label{s24}
\frac{1}{m}+1>0
\end{equation}
is an essential condition. The combination of  the conditions (\ref{s23a}) and (\ref{s24}) shows that
\begin{equation}\label{s25}
0<\frac{1}{m}+1<\alpha
\end{equation}
Secondly, we assume that $\omega_{r}>0$ which leads to
\begin{equation}\label{s26}
\frac{1}{\alpha}(\frac{1}{m}+1)>1.
\end{equation}
The condition $H_2>0$ gives
\begin{equation}\label{s27}
2\alpha>\frac{1}{m}+1>0
\end{equation}
For $\alpha>0$, using (\ref{s26}) and (\ref{s27}) leads to
\begin{equation}\label{s27a}
2\alpha>\frac{1}{m}+1>\alpha
\end{equation}

The same strategy can be used to check the condition  $H_{3}>0$. First, we assume  $\omega_{t}<0$ which results in
\begin{equation}\label{s28}
\frac{1}{2\alpha}(\frac{1}{m}+1)<1
\end{equation}
so
\begin{equation}\label{s29}
H_3(r)=1-|\omega_t|=\frac{1}{2\alpha}(\frac{1}{m}+1).
\end{equation}
The condition $H_3>0$ gives
\begin{equation}\label{s29a}
0<\frac{1}{2\alpha}(\frac{1}{m}+1).
\end{equation}
Combination of (\ref{s28}) and (\ref{s29a}) leads to
\begin{equation}\label{s30}
0<(\frac{1}{m}+1)<2\alpha.
\end{equation}
In the second case, we assume  $\omega_{t}>0$   which leads to
\begin{equation}\label{s31}
\frac{1}{2}(\frac{1}{m}+1)>\alpha.
\end{equation}
 One can easely show that $H_3>0$ gives
 \begin{equation}\label{s32}
2>\frac{1}{2}(\frac{1}{m}+1)
\end{equation}
 so the combination of (\ref{s31}) and (\ref{s32}) leads to
 \begin{equation}\label{s33}
\alpha<\frac{1}{2}(\frac{1}{m}+1)<2\alpha.
\end{equation}

We summarize the results regarding the EC violations in Table \ref{Tab1}. The table shows that these violations explicitly depend on the condition $0 < 1 + 1/m$ for NEC and WEC. For the DEC and SEC, an additional limitation is imposed by the appearance of $\alpha$. For instance, setting $n=1$ yields $\alpha=1/3$. In this case, the condition
  \begin{equation}\label{s34}
\frac{1}{3}<\frac{1}{2}(\frac{1}{m}+1)<\frac{2}{3}
\end{equation}
ensures that all ECs are satisfied. For example, $m=-2$ satisfies condition (\ref{s33}), which, according to Table \ref{Tab1}, demonstrates that $\beta < 0$ is acceptable. One can easily find that $\omega_r=-2\omega_t=1/2$.




\begin{table*}[]
\begin{tabular}{|l|l|l|l|}
\hline
 & Main condition & $\beta>0$  & $\beta<0$  \\ \hline
$\rho>0$  & $\alpha=\frac{1}{2n+1}$   & $0<m<1$ & $m<-1$\\ \hline
  $H>0$ &$0<1+\frac{1}{m}$  & $0<m<1$&  $m<-1$  \\ \hline
$H_1>0$ &$0<1+\frac{1}{m}$  & $0<m<1$&  $m<-1$ \\ \hline
$H_2>0$ &$0<1+\frac{1}{m}<2\alpha$  & $0<m<1$&  $m<-1$ \\ \hline
$H_3>0$ &$0<1+\frac{1}{m}<4\alpha$  & $0<m<1$&  $m<-1$ \\ \hline
$H_4>0$ &$\alpha<1+\frac{1}{m}$  & $0<m<1$&  $m<-1$ \\ \hline
\end{tabular}
\caption{Summary of results for the ECs and their relationship with the parameters $\alpha$, $\beta$, and $m$. }\label{Tab1}
\end{table*}

\section{Concluding remarks}\label{sec4}

Despite its remarkable triumphs in astrophysics and cosmology, GR is understood within high-energy physics to be an effective field theory. Consequently, its framework must be extended to include  corrections in the ultraviolet limit, especially when probing the extreme conditions of spacetime singularities, black holes, and wormholes. Within standard GR, the construction of a traversable wormhole requires the violation of NEC, demanding the existence of "exotic matter" possessing negative energy density.
 Because no known macroscopic form of matter naturally exhibits this property in standard physics, researchers have extensively turned to modified theories of gravity as a natural framework to support wormhole geometries.

Among the most compelling extensions of GR are teleparallel and metric-affine gravity theories, which shift the geometric foundation of spacetime. While standard GR is formulated on a Riemannian manifold characterized uniquely by curvature (with vanishing torsion and non-metricity), modified geometries allow for spacetime to possess torsion, non-metricity, or both.  In this paper, the field equations of $f(Q,L_m)$ gravity are derived by varying the action. Both linear and non-linear $f(Q,L_m)$ models are investigated. We have shown that for the linear model, the structure of the field equations is identical to that of GR, with the only difference arising from the inclusion of the term $\frac{\alpha}{\beta}$, which can be interpreted as a shift in the effective gravitational coupling constant. Furthermore, based on previous studies \cite{foad3, SR2, SR1, foad4, fq-sara}, we conclude that solutions violating both the radial and lateral NEC simultaneously in GR can satisfy the ECs within the framework of $f(Q,L_m)$ gravity.

We have demonstrated that, under specific conditions for the parameters $\alpha$, $\beta$, and $m$, a power-law shape function ($b(r)=r^m$) can support non-exotic wormhole solutions within both linear and non-linear $f(Q,L_m)$ models for a linear EoS. For the non-linear model, the primary condition is that $0 < 1 + 1/m$, where the role of $\alpha$ imposes certain limitations on the value of $m$, as shown in Table \ref{Tab1}. By coupling non-metricity directly to matter, $f(Q,L_m)$ gravity introduces additional degrees of freedom that fundamentally alter standard gravitational dynamics. This interaction produces effective stresses capable of sustaining wormhole throats without requiring exotic matter. Consequently, $f(Q,L_m)$ wormholes offer critical insights into how geometry-matter coupling influences spacetime topology, ECs, and traversability. Ultimately, investigating wormholes through these geometric-coupling frameworks eliminates the need for hypothetical matter, anchoring traversability in the intrinsic fabric of spacetime and opening new avenues in modified gravity phenomenology.

\end{document}